\documentclass{article}
\usepackage{ijcai19}

\usepackage{times}
\usepackage{soul}
\usepackage{url}
\usepackage[hidelinks]{hyperref}
\usepackage[utf8]{inputenc}
\usepackage[small]{caption}
\usepackage{graphicx}
\usepackage{amsmath}
\usepackage{amsfonts}
\usepackage{booktabs}
\usepackage{pbox}
\usepackage{subcaption}
\usepackage{caption}
\usepackage{{tikz}}
\usepackage[ruled,longend]{algorithm2e}
\definecolor{mygreen}{rgb}{0,0.6,0}
\definecolor{mymauve}{rgb}{0.58,0,0.82} 

\usetikzlibrary{arrows,shapes, decorations.pathmorphing,backgrounds,positioning,calc}

\title{Identifying Illicit Accounts in Large Scale E-payment Networks - \\A Graph Representation Learning Approach}

\author{
Da Sun Handason Tam$^{1,2}$\footnote{Contact Author}\and
Wing Cheong Lau$^1$\and
Bin Hu$^2$\and\\
Qiu Fang Ying$^2$\and
Dah Ming Chiu$^1$\and
Hong Liu$^2$
\affiliations
$^1$The Chinese University of Hong Kong\\
$^2$Tencent Technology Co. Ltd, China\\
\emails
\{dstam, wclau\}@ie.cuhk.edu.hk,
\{glennbinhu, qiufangying\}@tencent.com,\\
\{dmchiu\}@ie.cuhk.edu.hk, 
\{wienerliu\}@tencent.com
}

\begin{document}

\maketitle
\begin{abstract}
Rapid and massive adoption of mobile/ online payment services have brought new 
challenges to the service providers as well as regulators in safeguarding 
the proper uses such services/ systems. In this paper, we leverage recent
advances in deep-neural-network-based graph representation learning to detect abnormal/ suspicious financial transactions 
in real-world e-payment networks. In particular, we propose an end-to-end Graph Convolution Network (GCN)-based algorithm
to learn the embeddings of the nodes and edges of a large-scale time-evolving graph.
In the context of e-payment transaction graphs, the resultant node and edge embeddings can 
effectively characterize the user-background as well as the financial transaction patterns of individual account holders.
As such, we can use the graph embedding results to drive downstream graph mining tasks such as node-classification 
to identify illicit accounts within the payment networks. 
Our algorithm outperforms state-of-the-art schemes including GraphSAGE, Gradient Boosting Decision Tree and Random Forest to deliver considerably higher accuracy (94.62\% and 86.98\% respectively)
in classifying user accounts within 2 practical e-payment transaction datasets.  
It also achieves outstanding accuracy (97.43\%) for another biomedical entity identification task while using only edge-related information.
\end{abstract}

\section{Introduction}
With the rapid adoption of mobile payment services by the e-commerce company, 
massive and dynamic transactions become susceptible to security risks. 
Given the potentially huge financial losses caused by such vulnerabilities, there is an urgent need to develop regulatory technology.
Existing systems use rule-based methods to first automatically screen and flag suspicious payment activities. 
The flagged cases are then passed to human subject matter experts for labour-intensive manual examination. 
Unfortunately, these rule-based methods often suffer from extremely high false positive rates, 
Pricewaterhouse Coopers \cite{pwc2010} revealed that 90\% - 95\% of all alerts generated by existing Anti-Money-Laundering (AML) software were false positives. 
Likewise, \cite{europol2018} reported that the precision in detecting criminal fund transfer was only about 1\%. 
On the other hand, there are also systems which employ data-driven approaches by applying supervised machine learning techniques including Support Vector Machine, simple Neural Networks \cite{paula2016deep}, Decision Tree \cite{sudhakar2016two}, Genetic Algorithms \cite{Alsed2012GeneticCA}, etc. 
While conventional data-driven ML-based approaches may sound more principled theoretically, their empirical results are still not satisfactory. 
Furthermore, most conventional ML-based algorithms require some unrealistic assumptions, 
e.g. behaviour of users are independent and identically distributed (i.i.d.).
Unfortunately. customers' behaviour/ decisions may affect each other and thus can be highly correlated.

Recently, an important research frontier for deep learning has been on the design of new approaches to handle graphs/ networked data which capture complex relations between different entities of interest. 
Applications of deep learning on graphs have already demonstrated promising, state-of-the-art performance in the analysis and behavioural prediction of biological networks \cite{Zitnik2017,Zitnik2018,gao2018edge2vec}, transportation networks \cite{yu2018spatio,geng2019spatiotemporal,chai2018bike}, 
citation networks \cite{hamilton2017inductive,kipf2017gcn}, as well as online social networks \cite{qiu2018deepinf}. 
Key industrial players such as IBM \cite{weber2018scalable} and Paypal \cite{paypal2019} have also indicated their interest and/ or conducted preliminary studies on using the network representation learning approach to identify fraudulent activities over different types of payment networks.

The task of identifying illicit accounts in an e-payment network can be formulated as a node classification problem.
To do so, we will leverage the idea of node embedding. 
The goal is to automatically learn a better latent representation/ embedding of a node (account holder) so that the representation can capture the graph structural information representing e-payment patterns/ relationship between different pairs/ groups. 
The learnt embeddings can then be used to drive different downstream predictive tasks such as node classification and regression/ link prediction. 
Those algorithms can be broadly divided into 3 different approaches, namely,  Matrix factorization (e.g. \cite{ahmed2013distributed}, \cite{cao2015grarep}, \cite{qiu2018network}), Random Walk (e.g., DeepWalk \cite{perozzi2014deepwalk}, Node2Vec \cite{grover2016node2vec}, Metapath2vec \cite{dong2017metapath2vec}, SDNE \cite{wang2016structural}, LINE \cite{tang2015line}) and Graph Convolutional Networks (GCN). 
Both the Matrix factorization and Random Walk approach try to learn the embeddings so that the nodes which are close to each other in the graph will have similar embeddings (e.g. in terms of cosine distance, square distance). 

On the other hand, the GCN approach has recently been shown to consistently outperform the matrix factorization and random-walk based approaches. 
Under the GCN approach, embeddings are generated by aggregating and transforming the neighbors’ embeddings in a recursive manner. 
This is different from the other 2 approaches, as it requires learning a function to aggregate the neighbourhood rather than learning a lookup table. 
Therefore, the learnt function can be applied to a new graph for generating node embeddings of unseen nodes, which in turn, can enable inductive learning and transfer learning.
In summary, this paper has made the following technical contributions:
\begin{itemize}
    \item Inspired by DeepSet \cite{zaheer2017deep}, we propose a new message passing mechanism that allows edge information to propagate into node representation (i.e. embedding).
    \item We design and implement a scalable realization of the proposed algorithm to analyze real-world graphs containing millions of nodes.
    \item We demonstrate the efficacy of the proposed algorithm by delivering state-of-the-art node classification accuracy for 3 large-scale graph datasets in practice.
\end{itemize}

The rest of this paper is organized as follows: 
We discuss our problem formulation and provide the technical design details of the proposed EdgeProp algorithm in Section \ref{problem formulation}.
In Section \ref{scalability}, we describe the strategies to scale up the algorithm to handle graphs with millions of nodes.
Section \ref{experiment} presents the experimental results.
Section \ref{related work} discusses the related work. The paper is concluded in Section \ref{conclusion}.

\section{Problem Formulation and Methodologies} 
\label{problem formulation}
Consider a directed graph $\mathcal G:=(\mathcal V, \mathcal E)$ where $\mathcal V$ is the set of vertices and 
$\mathcal E$ denotes the set of edges. 
Let $N:=|\mathcal V|$ denote the total number of vertices and 
$M:=|\mathcal E|$ denote the total number of edges. 
Let $\vec{x}_i\in\mathbb{R}^F, \forall i \in \mathcal V$ be the node features of node $i$. 
Let $\vec{e}_{ij}\in\mathbb{R}^{P}, \forall (i, j) \in \mathcal E$ be the edge features between node $i$ and node $j$.
If the graph is undirected, we can transform it into a directed graph by replacing all undirected edge between two nodes i and j with two directed edges: one from $i \rightarrow j$ and the other from $i \leftarrow j$.
We can construct the graph where $\mathcal V$ represents the users in the mobile payment platform, 
and $\mathcal E$ represents the set of transactions between different users.
In the semi-supervised node classification task, ground truth labels are only available for a subset of nodes, and our goal is to classify every nodes in the graph.
Please refer to Table \ref{tab:notation} for notations and symbols.

\subsection{Our Approach}
\label{methodologies}
Our GCN model should be capable of incorporating multi-edge features as all the transaction records between 2 users are in the form of multi-edges, one per transaction, between the user-pair. 
This brings up two issues: 
i) How to describe/ characterize an edge? 
ii) How to design the graph neural network architecture so that the node embeddings can incorporate the edge information? 
In this section, we will discuss some possible ways to tackle these 2 challenges. 

\subsubsection{Characterizing an Edge}
To tackle the first issue, namely how to describe/ characterize an edge, possible approaches include: 
(i) Using hand-crafted features (e.g. average transaction volume, total transaction counts, mean/ variance of the inter-arrival time, etc) that capture the aggregated statistics of the transactions between every node-pair over the period of interest. 
(ii) Characterizing the multi-dimensional time-series of transactions via their marginal distributions as well as the corresponding auto/cross-correlation functions and ; 
(iii) Using an LSTM-based module to encode/ transform each transaction time-series into a fixed-dimensional representation and then train the LSTM-based neural network together with the Graph Convolutional Network in an integrated, end-to-end manner. 
In this paper, we will focus on Approach (i).  We are currently working on (ii) and (iii).

\subsubsection{Edge Propagation - Incorporating Edge Attributes in the GCNs}
\begin{figure}
\centering
\begin{tikzpicture}
	\node[circle, draw, thick, label=left:$\vec{z}_d^{(k-1)}$] (d) {};
	\node[circle, draw, thick, above=4em of d, label=right:$\vec{z}_a^{(k-1)}$] (a) {};
	\node[circle, draw, thick, below left=of d, label=below:$\vec{z}_b^{(k-1)}$] (b) {};
	\node[circle, draw, thick, below right=of d, label=below:$\vec{z}_c^{(k-1)}$] (c) {};

	\draw[-stealth, blue, thick] (c.135) -- node[sloped, above, black] {$\vec{e}_{cd}$} (d.-45);
	\draw[-stealth, blue, thick] (a.270) -- node[sloped, above, black] {$\vec{e}_{ad}$} (d.90);
	\draw[-stealth, blue, thick] (b.30) -- node[sloped, above, black] {$\vec{e}_{bd}$} (d.225);
	\node[circle, draw, thick, right=10.1em of d, opacity=0.8, label=left:$\vec{z}_d^{(k-1)}$] (d_new) {};
	\coordinate[right=10em of d] (A);
    \node[fill=blue!10, right=0.7em of d , style={single arrow, draw=none}, scale=0.8] {Concatentation};

	\node[circle, draw, thick, above=4em of d_new, label=right:$\vec{z}_a^{(k-1)} || \vec{e}_{ad}$] (j) {};
	\node[circle, draw, thick, below left=of d_new, label=below:$\vec{z}_b^{(k-1)} || \vec{e}_{bd}$] (k) {};
    \node[circle, draw, thick, below right=of d_new, label=below:$\vec{z}_c^{(k-1)} || \vec{e}_{cd}$] (l) {};
    
	\draw[-stealth, blue, thick] (l.135) -- node[sloped, above, black] {} (d_new.-45);
	\draw[-stealth, blue, thick] (j.270) -- node[sloped, above, black] {} (d_new.90);
	\draw[-stealth, blue, thick] (k.30) -- node[sloped, above, black] {} (d_new.225);
	\node[circle, draw, very thick, right=6em of d_new, opacity=1, label=right:$\vec{z}_d^{(k)}$] (d_update) {};
\draw[-stealth, mymauve, opacity=0.5, ultra thick] (d_new.0) -- node[black, above, opacity=1.0, scale=0.8] {\begin{tabular}{c}Aggregation \\(Equation~\ref{eq:sum_aggregator} and \ref{eq:embedding}) \end{tabular}} (d_update.180);

\end{tikzpicture}
\caption{Edge propagation}
\label{fig:edge prop}
\end{figure}
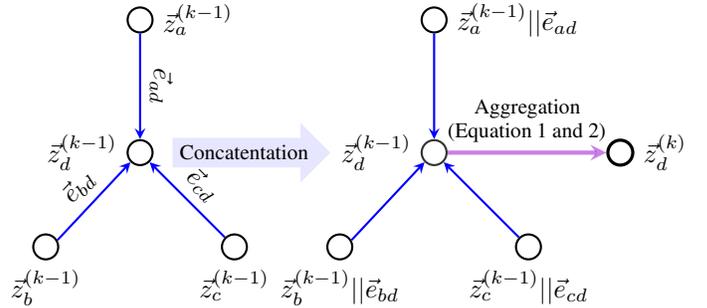
To tackle the second issue, namely, how to incorporate the rich set of edge information during the graph representation learning process, 
we propose an Edge Propagation mechanism that allows node embeddings to incorporate the edge embeddings based on the following idea inspired by DeepSet \cite{zaheer2017deep}: 
For any node $i$, we consider the set of neighbors as a multi-set. 
As \cite{xu2018how} has proven that any multiset function $f: \mathcal X \mapsto \mathbb R^n$ operating on countable set $X \subset \mathcal X$, 
there exists a function $\phi:\mathcal X \mapsto \mathbb R^n$ and a function $\rho:\mathbb R^n \mapsto \mathbb R^n$ so that $f(X)$ can be written in the form of $\rho(\sum_{x \in X}\phi(x))$.
To apply it to the message-passing paradigm, we can treat node $i$'s neighbor node embeddings and the associated edges as a multiset $\{(\vec{x}_j || \vec{e}_{ji}) \in (\mathbb R^F \times \mathbb R^P), \forall j \in \mathcal N_{\mathcal I}(i)\}$. 
We can construct the aggregation function to be $\rho_{\theta_2}(\sum_{j \in \mathcal N_{\mathcal I}(i)}\phi_{\theta_1}(\vec{x}_j || \vec{e}_{ji}))$, where $\rho_{\theta_2}$ and $\phi_{\theta_1}$ are multilayer perceptron with trainable parameters $\theta_2$ and $\theta_1$ respectively. 
In other words, we will concatenate the edge embeddings to the corresponding neighbour’s node embeddings and then perform the aggregation using the DeepSet architecture. 
Figure~\ref{fig:edge prop} depicts the workings of this approach.
Formally, the neighborhoods' representations can be defined recursively as follows:
\begin{align}
    \label{eq:sum_aggregator}
    \vec{z}_{\mathcal N_{\mathcal I}(v)}^{(k)} := &\sum_{u \in \mathcal N_{\mathcal I}(v)}\phi^{(k)}(\vec{z}_u^{k-1} || \vec{e}_{uv})\\
    \label{eq:embedding}
    \vec{z}_{v}^{(k)} := &\rho^{(k)}(\vec{z}_{v}^{(k-1)} || \vec{z}_{\mathcal N_{\mathcal I}(v)}^{(k)})
\end{align}

Although \cite{xu2018how} has shown that the sum aggregator is more expressive than the mean aggregator, in our experiments, the mean aggregator consistently performs better over various datasets. 
Therefore, in all our experiments (Section \ref{experiment}), we will use the mean aggregator to replace the summation operation in the R.H.S. of Equation~\ref{eq:sum_aggregator} even though this may introduce an inductive bias. As such, we have:
\begin{align}
    \label{eq:mean_aggregator}
    \vec{z}_{\mathcal N_{\mathcal I}(v)}^{(k)} := &\frac{1}{|\mathcal N_{\mathcal I}(v)|}\sum_{u \in \mathcal N_{\mathcal I}(v)}\phi^{(k)}(\vec{z}_u^{k-1} || \vec{e}_{uv})
\end{align}
Note that Equation~\ref{eq:mean_aggregator} is similar to \cite{hamilton2017inductive} except that we have included the edge embeddings and the new function $\phi(.)$ before the aggregation.

\begin{figure}
    \centering
    \begin{tikzpicture}
        \node[circle, draw, thick, label=left:$A$] (a) {};
        \node[circle, draw, thick, right=of a, label=right:$B$] (b) {};
        \path [->,bend left=40] (a) edge node[above] {$\vec{e}_{AB}$} (b);
        \path [->,bend left=40] (b) edge node[below] {$\vec{e}_{BA}$} (a);
    
        \node[circle, draw, thick, right=9em of b, opacity=0.8, label=left:$A$] (new_a) {};

        \node[fill=blue!10, right=1.7em of b , style={single arrow, draw=none}, scale=0.8] {Augmentation};
        \node[circle, draw, thick, right=of new_a, label=right:$B$] (new_b) {};
        \path [->,bend left=40] (new_a) edge node[above] {$\vec{e}_{AB}||\vec{e}_{BA}$} (new_b);
        \path [->,bend left=40] (new_b) edge node[below] {$\vec{e}_{BA}||\vec{e}_{AB}$} (new_a);
    \end{tikzpicture}
    \begin{tikzpicture}
        \node[circle, draw, thick, label=left:$A$] (a) {};
        \node[circle, draw, thick, right=of a, label=right:$B$] (b) {};
        \path [->] (a) edge node[above] {$\vec{e}_{AB}$} (b);
    
        \node[circle, draw, thick, right=9em of b, opacity=0.8, label=left:$A$] (new_a) {};

        \node[fill=blue!10, right=1.7em of b , style={single arrow, draw=none}, scale=0.8] {Augmentation};
        \node[circle, draw, thick, right=of new_a, label=right:$B$] (new_b) {};
        \path [->,bend left=40] (new_a) edge node[above] {$\vec{e}_{AB}||\vec{0}$} (new_b);
        \path [->,bend left=40] (new_b) edge node[below] {$\vec{0}||\vec{e}_{AB}$} (new_a);
    \end{tikzpicture}
    \caption{Augmenting edge features. This allows our method to capture the outgoing edge features}
    \label{fig:directed graph}
\end{figure}
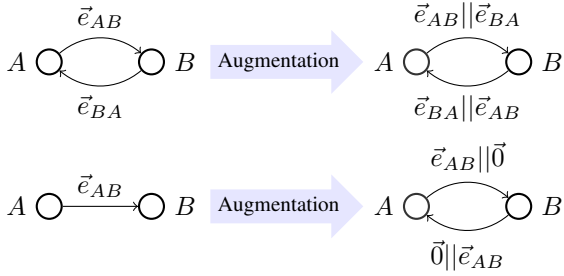
So far, we have only considered the aggregation of incoming neighbors (see Equation~\ref{eq:sum_aggregator} and Equation~\ref{eq:mean_aggregator}).
This is not optimal as it ignores all the outgoing neighbors. 
For example the total amount of payments that the person sends out can be important in identifying whether he/she has an extravagant lifestyle.
In order to capture the information of the outgoing edges, we can concatenate the edge features with the edge features in the reverse direction during the preprocessing step. If there is no edge in the reverse direction, we can zero pad the edge features so that the dimension of the edge feature-vector can remain the same. Figure~\ref{fig:directed graph} illustrates the edge augmentation process. After this augmentation, we can apply Equation~\ref{eq:sum_aggregator} or \ref{eq:mean_aggregator} to the representation, i.e. embeddings, of the neighboring nodes.

\begin{table}
\centering
\resizebox{\columnwidth}{!}{%
    \begin{tabular}{cl}
        \toprule
        Notation     & Description \\
        \midrule
        $deg(i)$     & Degree of node $i$     \\
        $\vec{z}_i^{(k)}$  & Node embeddings of node $i$ at layer $k$   \\
        $\vec{h}_i^{(k)}$  & Historical activation of node $i$ at layer $k$   \\
        $\vec{x}_i$  & The node features associated with node i, a F-dimensional vector\\
        $\vec{e}_{ij}$  & The edge features associated with the edge (i $ \rightarrow$ j), a P-dimensional vector\\
        $y_{i}$  & The ground truth label of node $i$\\
        $\| X \|_2$  & The L2 norm of vector X          \\
        $\mathcal V$ & Set of vertices \\ 
        $\mathcal V_{\mathcal L}$ & Set of vertices that contains label\\ 
        $\mathcal V_{\mathcal B}$ & Set of vertices in the mini-batch $\mathcal B$\\ 
        $N$ & Number of vertices, $|\mathcal V|$\\
        $\mathcal E$ & Set of multi-edges\\ 
        $M$ & Number of edges, $|\mathcal E|$\\
        $\mathcal N_{\mathcal I}(i)$       & The one-hop incoming neighbours of node i \\
        $\hat{\mathcal N_{\mathcal I}}(i)$       & The one-hop sampled incoming neighbours of node i \\
        $L$       & Number of layers \\
        $MLP$ & Mullti-layer perceptron \\
        $(\vec{x}||\vec{y})$ & The concatenation of $\vec{x}$ and $\vec{y}$\\
        \bottomrule
    \end{tabular}%
}
\caption{Mathematical notations}
\label{tab:notation}
\end{table}

\section{Scaling up for Large Graphs}
\label{scalability}
To scale up the model for large graphs, we adopt the ideas of mini-batch training and neighbor sampling.
The idea of mini-batch training is to approximate the full gradient with a stochastic gradient, which enables faster training.
The idea of neighbor sampling is to reduce the receptive field size during training, which reduces the memory footprint as well as enabling faster training.

\subsection{Mini-Batch Training}
We use stochastic gradient descent to update the parameters, which requires calculating the gradient of the loss function:
\begin{align}
\label{eq:full_grad}
\nabla \mathcal L =& \frac{1}{|\mathcal V_{\mathcal L}|} \sum_{v \in \mathcal V_{\mathcal L}} \nabla f(y_v, \vec{z}_v^{(k)})
\end{align}
where $f(y_v, \vec{z}_v^{(k)})$ is the loss function. 
In the multi-class node classification task, $f(y_v, \vec{z}_v^{(k)})$ is the cross-entropy loss. 
Calculating the full gradient as in Equation~\ref{eq:full_grad} is computationally expensive as it requires the calculation of the gradient of the losses for all vertices. 
To scale up this procedure for large graphs, we approximate the full-batch gradient by the following stochastic mini-batch gradient:
\begin{align}
\nabla \mathcal L =& \frac{1}{|\mathcal V_{\mathcal B}|} \sum_{v \in \mathcal V_{\mathcal B}} \nabla f(y_v, \vec{z}_v^{(k)})
\end{align}
Note that the mini-batch gradient is biased due to the non-linear operations over the neighboring samples \cite{chen2017stochastic}. As such, this approach may violate the convergence assumption of stochastic gradient descent. 
However, it is a consistent estimator and if we apply the control variate technique (Section \ref{ns}), 
the control variate gradient is asymptotically unbiased, and it can be shown that the convergence of SGD can still be guaranteed \cite{chen2017stochastic}. 
Alg. \ref{alg:edgeprop} depicts the algorithm of the mini-batch embedding generation process using the neighbor sampling technique.

\subsection{Neighborhood Sampling}
\label{ns}
In graph neural networks training, the mini-batch gradient is still expensive to compute as the receptive field size grows exponentially with the number of layers.
In the transaction dataset from Tencent, the 2-hop neighborhood of 20000 sampled nodes already results in a graph with 6.5 million nodes.
Towards this end, we will sample the neighborhood rather than using all the neighbors of each target node during the training phase.
The idea of neighborhood sampling/ vertex sampling has been proposed by GraphSAGE \cite{hamilton2017inductive}, FastGCN \cite{chen2018fastgcn} and StochasticGCN \cite{chen2017stochastic}.
To reduce the variance of the neighbor sampling estimator, we leverage the idea of control variate proposed by \cite{chen2017stochastic}, which works by maintaining a history of activations $h_i^{(l)}\; \forall i \in \mathcal V, l \in (1, ..., L)$ of various layers. 

Let $\Delta \phi^{(k)}(\vec{z}_v^{(k)} || \vec{e}_{uv})$ be $\phi^{(k)}(\vec{z}_v^{(k)} || \vec{e}_{uv}) - h_v^{(k)}$. With neighbor sampling and control variate, our new node embedding estimator becomes:

\begin{align}
    \vec{z}_{\mathcal N_{\mathcal I}(v)}^{(k)} \approx & \frac{1}{|\hat{\mathcal N_{\mathcal I}}(v)|} \sum_{u \in \hat{\mathcal N_{\mathcal I}}(v)}\Delta \phi^{(k-1)}(\vec{z}_u^{(k-1)}||\vec{e}_{uv}) \nonumber\\
    + & \frac{1}{|\mathcal N_{\mathcal I}(v)|} \sum_{u \in \mathcal N_{\mathcal I}(v)} h_u^{(k-1)}\\
    \vec{z}_v^{(k)} \approx & \rho^{(k)}(\vec{z}_v^{(k-1)} || \vec{z}_{\mathcal N_{\mathcal I}(v)}^{(k)})
\end{align}

\begin{algorithm}[tb]
    \SetKwInOut{Input}{input}
    \SetKwInOut{Output}{output}
    
    \Input{
        $Graph \; \mathcal G(\mathcal V, \mathcal E)$; \\
        input features ${\vec{x}_v}, \forall v \in \mathcal B$; \\
        edge features ${\vec{e}_{ij} }, \forall (i, j) \in \mathcal B$; \\
        depth $K$; \\
        differentiable functions $\rho, \phi$ (e.g. a MLP); \\
        neighborhood sampling function $\mathcal N_k: v \mapsto 2^{N}$
    }
    \Output{Vector representations $\vec{z_v}, \forall v \in \mathcal V$ }
    $\mathcal B^K \leftarrow \mathcal B$\;
    \For{$k=K...1$}{
        $B^{k-1} \leftarrow \mathcal B^k$\;
        \For{$u \in \mathcal B^k$}{
            $B^{(k-1)} \leftarrow B^{(k-1)} \cup \mathcal N_k(u)$
        }
    }

    $\vec{z}_v^0 \leftarrow \vec{x}_v, \forall v \in \mathcal B^0$\;
    \For{$k=1...K$}{
        \For{$v \in \mathcal B^k$}{
            $\vec{z}_{\mathcal N(v)}^{(k)} \leftarrow \sum_{u \in \mathcal N(v)}\phi^{(k)}(\vec{z}_u^{(k-1)} || \vec{e}_{uv})$\;
            $\vec{z}_v^{(k)} \leftarrow \rho^{(k)}(\vec{z}_v^{k-1} || \vec{z}_{\mathcal N(v)}^{(k)})$\;
        }
    }
    \KwRet{$\{\vec{z}_v^{(k)} \forall v \in \mathcal V\}$}\;
    \caption{Embedding Generation (i.e. forward propagation) Algorithm}
    \label{alg:edgeprop}
\end{algorithm}

\section{Experimental Results}
\label{experiment}
\subsection{Setup}
In this section, we evaluate the performance of our proposed algorithms using 3 network datasets in practice. 
We seek to answer 2 questions: 
i) Are edge features useful for node classification?
ii) Does EdgeProp improve the overall performance of the node classification task?
The EdgeProp algorithm is implemented using Pytorch \cite{paszke2017automatic}, using the DGL \cite{wang2019dgl} Python package. 
We use the ADAM \cite{kingma2014adam} optimizer with learning rate of $2e^{-4}$.
Rectified linear unit \cite{nair2010rectified} is employed as the non-linear activation function across all layers.
We apply both mini-batch sampling with 32 batch size and neighborhood sampling (uniform distribution with sample size 10) to speed up the training and use the control variate method \cite{chen2017stochastic} to reduce the variance of the neighbor sampling estimator.
The mean aggregator (Equation~\ref{eq:mean_aggregator} and \ref{eq:embedding}) is used for generating the representation of each node.
We use multilayer perceptron with a single hidden layer for both $\rho(.)$ and $\phi(.)$.
The hidden layer size, as well as the embedding size, is set to be 32.
Early-Stopping with windows size 100 is used to determine the stopping time. 
For each datasets, we split the set of vertices with ground label $\mathcal V_{\mathcal L}$ into train set (70\%), validation set (10\%) and test set (20\%).
We ran all experiments in a single machine with Intel(R) Xeon(R) CPU (E5-2680 v4 @ 2.40GHz), 
and 128 GByte of RAM, and a NVIDIA(R) Tesla(R) M40 24GB of RAM. 
Node features, edge features and the graph are stored in main memory.

We compare our results against 6 baseline: 
Logistic Regression, Random Forest, Gradient Boosting Decision Tree (GBDT), DeepWalk, Line and GraphSAGE.
\begin{itemize}
\item For Logistic Regression, Random Forest and Gradient Boosted Tree, they only use the node features as input (i.e. ignoring the graph structure).
\item For DeepWalk and Line, we concatenate the learnt embeddings with the corresponding node features. The embedding size is set to be 128. 

\item For GraphSAGE, we use the mean aggregation method.
\end{itemize}

We report the Accuracy, Macro Precision, Macro Recall and Macro F1 score on the test set.

\subsection{Ethereum Dataset}

\begin{table}
    \centering
    \resizebox{\columnwidth}{!}{%
        \begin{tabular}{cl}  
        \toprule
        Categories  & Description \\
        \midrule
        ICO & ICO wallet used for crowdsale/presale \\
        Trader & Traders who are active in the past two years \\
        Converter & Automatic token conversion wallet \\
        Exchange & Hot wallet of exchange platforms \\
        Mining & Accounts of mining pools \\
        Gambling & Popular gambling contract accounts \\
        Scam& Scam (e.g. Phishing scam) accounts reported by users\\
        \bottomrule
        \end{tabular}
    }
    \caption{Details of the node categories (labels) in our Ethereum Dataset}
    \label{tab:Eth classes}
\end{table}
Ethereum is currently the second largest blockchain platform. 
We have collected a dataset of Ethereum transactions for the entire year of 2018 using Google BigQuery.
To train our node classification model using supervised learning, 
we obtain 1024 ground-truth node labels of official public accounts from Etherscan \cite{team2017etherscan}. 
We then construct the graph by considering all the addresses that are within the one-hop neighborhood of each target node. 
Since an actual Ethereum user can create as many accounts as he/she wishes, 
we observe that most of the addresses (accounts) are only involved in one single transaction during the entire year of 2018. 
We, therefore, filter out the inactive addresses by removing all the nodes which are involved in fewer than 3 transactions counts or the total transaction value is less than 0.1 eth during 2018. 
The result is a network of 2.18 million nodes and 3.64 million edges and 445 ground truth labels. 
15 Node features (e.g. balance, in degree, total count of the smart contracts creation, number of days the account have activity records, e.t.c.) and 10 edge features (e.g. total transaction value, average transaction value, variance of the transaction value, variance of the inter-arrival time, etc) are extracted from the transaction records. 

We then use EdgeProp to predict the type/ category of the user associated with each Ethereum address. This is a multi-class classification problem as there are 7 types of ground-truth labels (see table \ref{tab:Eth classes} for details) among the training data. 
Table~\ref{tab:Ethereum result} shows the experimental results. Note that EdgeProp with node features (abbreviated as w/ nft.) outperforms all the other baseline schemes in terms of accuracy. In particular, EdgeProp (w/ nft.) achieves node classification accuracy of 94.62\%. 
However, it does not achieve the best macro-precision, macro-recall and macro-f1 score. This may be due to the limited number of ground truth labels available for training the GCN. 
It is noteworthy that EdgeProp can still achieve competitive results (80.65\% accuracy) even without considering any node attributes.
The fact that DeepWalk, Line and Logistic Regression produce the exact same predictions on the test set indicates that the learnt embeddings of DeepWalk and Line do not provide additional information on top of the raw node features in predicting the Ethereum node classes.
Notice that we do not include Edge2Vec in the comparison as it requires discrete edge types. In contrast, the edge attributes are multi-dimensional continuous-valued features in the Ethereum dataset.
\begin{table}
    \centering
    \resizebox{\columnwidth}{!}{%
        \begin{tabular}{ccccc}  
        \toprule
        Algorithm  & Accuracy & Precision     & recall & f1 measure\\
        \midrule
        DeepWalk   & 0.9101 & $\mathbf{0.6809}$ & 0.6254 & 0.6520 \\
        Line       & 0.9101 & $\mathbf{0.6809}$ & 0.6254 & 0.6520 \\
        Logistic Regression & 0.9101  & $\mathbf{0.6809}$ & 0.6254 & 0.6520 \\
        Random Forest    & 0.9438  & 0.6583 & 0.6838 & 0.6708 \\
        GBDT             & 0.9213  & 0.6397 & $\mathbf{0.7716}$ & $\mathbf{0.6995}$\\
        GraphSAGE        & 0.8602 & 0.5171 & 0.4491 & 0.4807 \\
        EdgeProp (w/ nft. + incoming)  & $\mathbf{0.9462}$   & 0.6426  & 0.7078       &  0.6736 \\
        EdgeProp (w/ nft. + directed)  & $\mathbf{0.9462}$   & 0.6426  & 0.7078       &  0.6736 \\
        \bottomrule
        \end{tabular}
    }
    \caption{Results on Ethereum dataset}
    \label{tab:Ethereum result}
\end{table}

\subsection{Mobile Payment Transactions Dataset}
Identifying illicit users who conduct gambling activities via mobile payment transactions is one of the most critical challenges for the system' security and reliability. 
In particular, we collect a dataset by randomly sampling 10k users with a ground-truth label of ``gambler'', and sampling an equal number of normal users.
We then construct the graph by including all the users within the two-hop neighbourhood of each targeting node as well as the transactions between them in the observation window of one month. 
After filtering out edges with less than two transaction count or low transaction amount, the resultant graph has around 6.49 million nodes and 33.63 million directed edges. 
61 node features (such as node profile information, e.g., age, gender, education, etc) identifying the users and 10 edge features (transaction amount, count, timestamp and their variants) identifying the transaction behavior between user-pairs are extracted.
To protect the samples' privacy, all user IDs are anonymized and all transactions' amount and frequency are normalized.

In the training process, we use neighborhood sampling size of 10, and only consider layer size = 1.
The training process consumes 60GB of CPU RAM and 12GB of RAM of the GPU.
Table~\ref{tab:wechat result} compares the performance of EdgeProp against that of the baseline schemes. In terms of classification accuracy, EdgeProp shows an improvement of more than 5\% (i.e. 86.98\% vs. 81.10\%) when comparing to GraphSAGE. Notice that EdgeProp with edge augmentation (abbreviated as directed) can still obtain competitive results ($83.65\%$ accuracy) even without using any node features (abbreviated as w/o nft.).

\begin{table}
    \centering
    \resizebox{\columnwidth}{!}{%
        \begin{tabular}{ccccc}
        \toprule
        Algorithm        & Accuracy & Precision & recall & f1 measure\\
        \midrule
        DeepWalk         & 0.6847  & 0.6648 & 0.6286 & 0.6462\\
        Line             & 0.6841  & 0.6636 & 0.6293 & 0.6460\\
        Logistic Regression & 0.6822  & 0.6253 & 0.6621 & 0.6432 \\
        Random Forest    & 0.7767  & 0.7474 & 0.7490 & 0.7482 \\
        GBDT             & 0.7634  & 0.7284 & 0.7483 & 0.7382\\
        GraphSAGE        & 0.8110 & 0.8138 & 0.8053 & 0.8095 \\
        EdgeProp (w/o nft. + incoming) & $0.8303$   & $0.8305$ & $0.8330$ & $0.8317$ \\
        EdgeProp (w/o nft. + directed) & $0.8365$   & $0.8375$ & $0.8399$ & $0.8387$ \\
        EdgeProp (w/ nft. + incoming)   & $0.8639$   & $0.8633$ & $0.8622$ & $0.8627$ \\
        EdgeProp (w/ nft. + directed)   & $\mathbf{0.8698}$   & $\mathbf{0.8686}$ & $\mathbf{0.8696}$ & $\mathbf{0.8690}$ \\
        \bottomrule
        \end{tabular}
    }
    \caption{Results on Tencent transaction dataset}
    \label{tab:wechat result}
\end{table}

\subsection{Drug Discovery Dataset}
We also consider the task of biomedical entity identification using the  Chem2Bio2RDF drug discovery dataset \cite{chen2010chem2bio2rdf} which contains a heterogeneous graph with 10 different types of nodes and 12 different types of edges. 
The graph has a total of 295,911 nodes and 727,997 edges.
For the biomedical entity identification task (node classification), only the edge type and the graph topology information are used as input (i.e. no node features).
For details, please refer to the Edge2Vec \cite{gao2018edge2vec} paper.
In short, EdgeProp achieves excellent accuracy of 97.43\% for the task of classifying different biomedical entities such as genes, tissue and disease within the dataset.

\section{Related Work}
\label{related work}
Despite the empirical success of GCNs, 
there have been very limited work that considers a graph with edge attributes. 
Most GCN models in the literature do not accept edge-attributes as input. Even for those few which support edge-attributes, all but one of them restrict their consideration to heterogeneous graphs with limited number of types of edges.
In particular, Decagon \cite{Zitnik2018} leverages the Relational GCN \cite{schlichtkrull2018modeling} model and tries to learn a different set of parameters for each edge type (e.g. Gastroinestinal bleed side effect, Bradycardia side effect). 
It uses the learned embeddings to predict side effect of drug combinations. 
Edge2vec \cite{gao2018edge2vec} focuses on identifying biological entities such as genes, proteins, drugs, diseases problem, 
it learns a transition probability matrix among the edge types by the EM algorithm. 
Then, a random walker is used to extract many sequences of nodes based on the learned transition probability. 
They use the approach similar to deep walk to learn the embeddings of each node and used it for classification tasks. 
Note that both Decagon and Edge2vec can only able to deal with graphs with edge-types. 
Metapath2vec \cite{dong2017metapath2vec} leverages DeepWalk \cite{perozzi2014deepwalk} and a pre-defined meta-path (a sequence of edge type). 
The random walker is restricted to follow the meta-path. 
Similar to DeepWalk, they leverage the skip-gram model to learn the node embeddings.
This method can be extended to graph with edge-types by incorporating edge type into the meta-path.
However, multi-dimensional edge features cannot be incorporated directly.

Since GAT has an edge-wise mechanism, they are naturally allowed to support any kind of edge features. 
In the simplest case, we can feed the edge features as the extra input parameters for the attention. 
EGNN \cite{liyu2018exploiting}, extends this idea by treating each edge feature separately and uses the attention function which takes the node embeddings, as well as the edge features as input, 
and produces the attention coefficient. 
Weighted sum with the attention coefficient is being applied to the neighbors' features to perform the aggregation.
Although this method utilizes the edge features, the edge features are being interpreted as connectivities/ strengths between node pairs.
On the other hand, in our proposed method, we treat them as a multi-dimensional vector, we allow complex interaction between the edge features and the node features by introducing $\phi(.)$ and $\rho(.)$.

\section{Conclusion}
\label{conclusion}
In this work, we have presented the EdgeProp - a new message passing mechanism which allows multi-dimensional continuous edge features propagating into node embeddings when performing node classification tasks.
This method is designed to address the common issues of the state-of-the-art GCN models where edge features are either ignored or poorly supported.
Our method is scalable and can be integrated to different graph neural network models.
Empirical results show that EdgeProp outperforms current state-of-the-art models such as GraphSAGE and GBDT. 
It also shows that edge features are important inputs for the node classification task.
Futuer work will examine different variants of message passing mechanisms with richer set/ types of edge features. Currently, we are also extending EdgeProp so that the edge embeddings of the graph of interest can be learnt directly from the multi-dimensional time-series of feature-rich payment transactions.


\appendix

\bibliographystyle{named}

\end{document}